# Crystallographic Reconstruction Driven Modified Mechanical Properties in Anisotropic Rhenium Disulfides

Jung Hwa Kim[1], Xinyue Dai[1], Feng Ding[1,2*], and Zonghoon Lee[1,2*]

[1]Center for Multidimensional Carbon Materials, Institute for Basic Science (IBS), Ulsan 44919, Republic of Korea
[2]Department of Materials Science and Engineering, Ulsan National Institute of Science and Technology (UNIST), Ulsan 44919, Republic of Korea

*Corresponding author E-mail: zhlee@unist.ac.kr (Z.L.); f.ding@unist.ac.kr (F.D.)

Atomic-scale investigation on mechanical behaviors is highly necessary to fully understand the fracture mechanics especially of brittle materials, which are determined by atomic-scale phenomena (*e.g.*, lattice trapping). Here, exfoliated anisotropic rhenium disulfide ($ReS_2$) flakes are used to investigate atomic-scale crack propagation depending on the propagation directions. While the conventional strain-stress curves exhibit a strong anisotropy depending on the cleavage direction of $ReS_2$, but our experimental results show a reduced cleavage anisotropy due to the lattice reconstruction in [100] cracking with high resistance to fracture. In other words, [010] and [110] cracks with low barriers to cleavage exhibit the ultimate sharpness of the crack tip without plastic deformation, whereas [100] cracks drive lattice rotation on one side of the crack, leading to a non-flat grain boundary formation. Finally, crystallographic reconstruction associated with the high lattice randomness of two-dimensional materials drives to a modified cleavage tendency, further indicating the importance of atomic-scale studies for a complete understanding of the mechanics.

To date, atomic-scale investigations of fractures have been consistently considered as essential to fully understand the various phenomena of mechanical behaviors. For example, cracks in brittle materials involve atomically sharp crack tips and their propagation is continuing by breaking the variously oriented interatomic bonds at the moving crack front, one at a time (*1, 2*). Moreover, lattice trapping or atomic reconstruction in the vicinity of the crack tip results in high fracture resistance, beyond Griffith's criterion providing a well-established description for linear elastic fracture (*3, 4*). In this way, the fracture mechanics are determined entirely by mechanisms on the atomic scale. Therefore, atomic-scale studies on fracture could play a key role in uncovering many of the remaining delicate issues (*e.g.*, the origin of fracture instabilities (*1, 5, 6*)). However, previous studies have been limited to numerical calculations instead of experimental atomic-scale crack propagation investigations (*1, 3-5*), and even calculations have been confined to isotropic materials, so few reports dealt with anisotropic materials (*7*).

Two-dimensional (2D) materials, which have been successfully isolated since 2004 (*8*), are considered as ideal materials for a comprehensive atomic-scale understanding of mechanical properties due to their inherent thinness and intuitive interpretation (*9*). Recently, *in situ* transmission electron microscopy (TEM) studies have revealed that the cracks in isotropic 2D materials (*e.g.*, graphene, hexagonal transition metal dichalcogenides) propagate along the energetically favorable zigzag edges (*10, 11*) with atomically sharp crack tips (*12*). Pure brittleness has been reported along the basal surface of graphene and hexagonal-molybdenite, but atomic-scale investigations using TEM have shown the emerging of dislocations and atomic reconstructions in the vicinity of the crack tips (*12-15*), beyond previous knowledge of fracture in bulk brittle materials. Moreover, the anisotropic fracture dynamics in monolayer tungsten disulfide depending on the types of elements were observed using *in situ* TEM, which could not be demonstrated by conventional fracture theory (*16*). By the way, in-plane anisotropic materials (*e.g.*, $ReS_2$, $WTe_2$) with low-symmetric atomic structures lead to anisotropic electronic, optical and especially mechanical behaviors (*17-22*). Therefore, the establishment of a relationship between intrinsic structure and mechanical property in 2D anisotropic materials is strongly required to enrich the fracture mechanics in systems of reduced dimensionality and anisotropic structures, but atomic-scale studies on the fracture mechanics of 2D anisotropic materials is still lacking.

Herein, we investigate the crack propagation of rhenium disulfide ($ReS_2$), a representative in-plane anisotropic 2D material, on an atomic scale. The exfoliated flakes predominantly have an elongated shape along the Re-chain direction ([010] direction), which could be explained using the low ultimate tensile strength in the uniaxial tensile stress-strain (SS) relationship. Other frequently observed (110) cleavage surfaces can also be described in the same way. Both show the ultra-sharpness of crack tip without atomic reconstruction and dislocation emission, indicative of ultra-brittleness of $ReS_2$ along these directions. However, despite the high resistance to fracture, the (100) cracking surfaces are often found with sharp and straight edges. Our atomic-scale investigation shows that when the cracking is along the [100] direction, the ultrathin $ReS_2$ near the crack tip accompanies a crystallographic rotation of one side of the crack and the other side remains, resulting in a formation of grain boundary (GB). Notably, GB generated by crack propagation exhibits even out-of-plane distortion, demonstrating the crack-induced 3D fluctuation in 2D materials. Therefore, we propose a crystallographic reconstruction mediated modified SS curve, indicative of high degree of lattice randomness of 2D materials. The reduction in the ultimate tensile strength and work function cut along the [100] cracks from atomic reconstruction results in a reduced anisotropy of the cleavage tendency in $ReS_2$, and it is beyond Griffith's description and previous understanding of cracking in isotropic materials.

Prior to the text, we first mention that the crack in our study began in two ways, as shown in the fig. S1; 1) crack occurring during sample transfer (*23*) and 2) residual tension-driven cracks in holes generated by electron beams (*12, 14*). In addition, since the synthesized $ReS_2$ flakes inevitably contain lots of lines defects (*e.g.*, GB, antiphase boundaries) created during synthesis (*24, 25*), mechanically exfoliated samples were exclusively used. Thus, it is possible to investigate the inherent fracture mechanics of $ReS_2$ occurring during cracking (*e.g.*, atomic reconstruction), excluding other intrinsic defect-related factors. Our TEM work was performed at room temperature with a sufficiently low beam intensity to minimize beam-induced lattice transformation. Detailed experimental conditions are described in **Methods**.

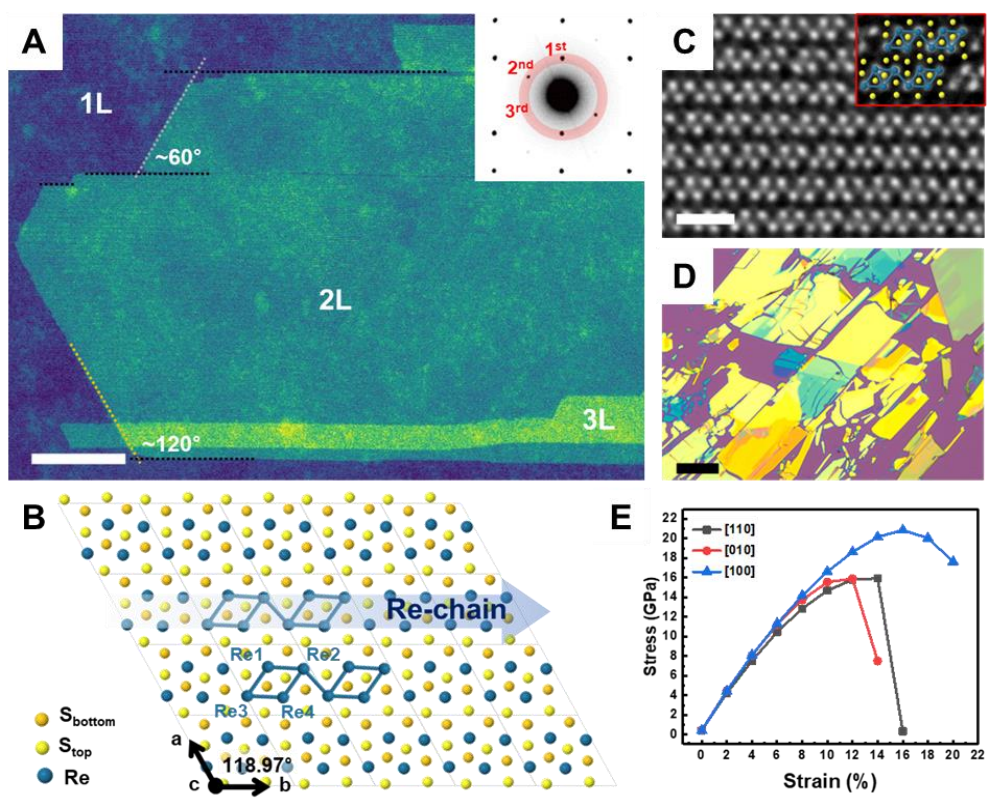

**Figure 1. Cleavage tendency of anisotropic ReS₂ flakes.** (**A**) STEM image at the low magnification of 1 to 3 layers of $ReS_2$ flake suspended on a conventional TEM grid. The inset is the corresponding SAED pattern, showing a reduced intensity counterclockwise at the red-shaded diffraction spots (for convenience, these are referred to as the 1st order diffraction spots), and consequently it can be concluded that the $ReS_2$ flake has an "A" structure. (see Supplementary Fig. 2) (**B**) Atomic modeling of single-layer $ReS_2$. Blue and orange/yellow balls indicate rhenium and sulfur (bottom/top) atoms, respectively. Re1 to Re4 atoms form diamond-shaped Re clusters, which are linked to form anisotropic Re-chains. (**C**) STEM Z-contrast image obtained from the single-layer $ReS_2$ in (A). The inset shows the structural model overlaid on the STEM image. (**D**) Optical image of typical $ReS_2$ flakes. Most samples have an elongated shape in one direction. (**E**) Calculated SS curves of single-layer $ReS_2$ system depending on the cleavage direction. The denoted directions are the crack propagation directions along [110] (black), [010] (red), and [100] (blue). Scale bar for (A), (C), and (D), 2 μm, 1 nm, and 10 μm, respectively.

Figure 1 depicts the anisotropic in-plane cleavage tendency of exfoliated $ReS_2$ flakes using the standard scotch tape method (*8*). The low-magnification STEM image in Fig. 1A shows bi- and tri-layer $ReS_2$ flakes, containing cleaved edges in typical three directions (indicated by black, white, and yellow-dashed lines). First it is necessary to fully identify the orientation and structural features of $ReS_2$ to determine the favorable cleavage direction. Compared to the widely studied group-VI TMDs (*e.g.*, $WS_2$, $MoS_2$), each rhenium atom in $ReS_2$ has one extra valence electron, forming a Re-Re bond. Therefore, distorted octahedral (1T′) structure of ReS$_2$ has the diamond-like shaped clusters of Re$_4$ units (Re1 to Re4) and these are linked by Re-Re bridge bonds to form quasi-one-dimensional Re-chains (Fig. 1B) (*24, 26*). The mechanical properties of ReS$_2$ are strongly correlated with these anisotropic structural features of Re-chains (*27*). In addition, it is noteworthy that the flipped structure of ReS$_2$ should be distinguished because it has one inversion center with triclinic symmetry (marked "A" and "A'" in fig. S2) (*28*). Here, we propose a SAED pattern analysis to fully determine the orientation of ReX$_2$ (X=S,Se). Using the direction of intensity decreasement of the 1$^{st}$ order SAED pattern (red-shaded region in inset of Fig. 1A) due to the different structure factors of the (100) and (110) planes, it can be concluded that the exfoliated flake (Fig. 1A) has an "A" configuration (fig. S2A,B). Additionally, the AR-STEM image (Fig. 1C) obtained from the single-layer region of Fig. 1A is in good agreement with the "A" atomic structure. Consequently, SAED analysis could be used to accurately determine the ReBX$_2$ orientation with the powerful advantages of time and accuracy.

Optical image with a large field-of-view (Fig. 1D) shows that the exfoliated ReS$_2$ flakes have an elongated shape mainly in one direction (*i.e.*, a greater probability of appearance of the edges for one direction), as in the previous report (*27*). With the combination of STEM and OM analysis, the exfoliated ReS$_2$ flakes were mainly cut along the (010) plane leaving the longest edges along the *b*-axis, while isotropic hexagonal TMDs (*29*) (*e.g.*, 2H phase MoS$_2$, 1T phase PtS$_2$) show a strong cleavage tendency along the family of (11$\bar{2}$0) lattice planes (*i.e.*, 6 ZZ directions).

To validate these anisotropic cleavage features in ReS$_2$, we performed numerical uniaxial tensile SS relationship of single-layer ReS$_2$ along six typical directions (Fig. 1E, fig. S3 and table S1). The indicated numbers (Fig. 1E) are the directions of crack propagation (fig. 3A). Want *et al.* reported that ReS$_2$ flakes have a nearly isotropic in-plane elastic response with Young's modulus $E_{ReS_2} \approx 190$ GPa (*27*). Therefore, the anisotropic cleavage tendency of ReS$_2$ is predicted to be closely related to the strong anisotropic ultimate tensile strength and critical strain in the SS curves of ReS$_2$. The lowest ultimate tensile strength and lowest critical strain along [010] direction of ReS$_2$ result in high probability to break along [010] directions, which is associated with the dissociation of the relatively weak Re-S bonds (solid red line in Fig. 1E). The [110] direction (solid black line in Fig. 1E), related to the large spacing between adjacent Re clusters (~0.3 nm), has a comparable ultimate tensile strength, but higher critical strain compared to the [010] crack, thus this is the edge most often seen after the [010] crack (fig. S4). In this way, numerical calculation data for [010] and [110] directions agree well with experimental cleavage phenomena (black and white-dotted edge in Fig. 1A, respectively). According to the calculation result, the crack propagation along [100] direction (solid blue line in Fig. 1E) has high resistance to breakage due to the highest ultimate strength, based on the strong metal-metal covalent bonding between Re atoms, but this direction is frequently observed to have sharp straight and long edges depicted in Fig. 1A (yellow line) and fig. S4. At this point, we propose that how sharp cleavage edges occur along the [100] direction might be related to mechanisms other than tensile strength, and to fully account for the cleavage tendency of anisotropic materials, we are going to investigate the atomic-scale cracking of ReS$_2$.

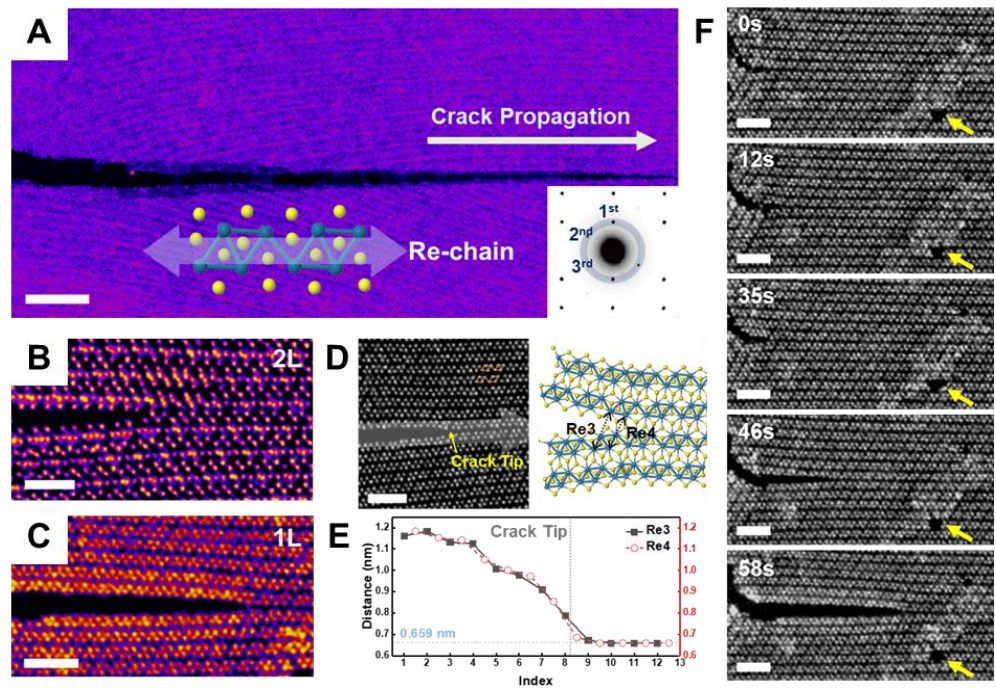

**Figure 2. Crack propagation along the [010] direction (parallel to the Re-chain direction).** (**A**) Large field-of-view STEM image of the [010] crack in bilayer ReS$_2$ and corresponding SAED pattern (inset). (**B,C**) Atomic-scale HAADF-STEM images of the crack tip zones in the bi-layer (B) and single-layer (C) of ReS$_2$. (**D**) Experimental STEM image and calculated atomic model near the crack tip propagating along the [010] direction. (**E**) Distance between Re atoms across the crack along the white arrow in (D). The indicated 0.659 nm is the distance between adjacent Re3 (Re4) atoms across the [010] direction in the pristine ReS$_2$. (**F**) Successive STEM images exhibiting the crack propagation along the [010] direction, showing no atomic reconstruction in the vicinity of the crack tip. Scale bar for (A), (B), and (C,D,F), 20 nm, 1 nm, and 2 nm, respectively.

Figure 2 presents the atomic-scale crack propagation along the [010] direction (*i.e.*, parallel to the Re-chain direction), which is the edges most often observed in exfoliated (*27, 30*) and CVD-synthesized (*25*) samples. As aforementioned, this direction has the lowest ultimate strength and critical strain due to the lack of strong Re–Re covalent bonds. Fig. 2A is a false-colored low-magnification STEM image of a crack along the Re-chain direction in bi-layer $ReS_2$. Our AR-STEM analysis of the crack (010) surface obtained near the crack tip in Fig. 2A shows an atomically sharp crack tip without lattice perturbation ahead of the crack tip (Fig. 2B). Similarly, in the case of [010] cracking in single-layer $ReS_2$, the crack proceeds only by decohesion of Re–S bonds with maintaining the perfect lattice crystallinity in front of the crack tip (Fig. 2C). It is excellent agreement with the calculated atomic model depicted in Fig. 2D. The atomic distance between adjacent Re3 (Re4) atoms across the crack begins to increase abruptly at the crack tip as in Fig. 2E obtained along the white arrow in Fig. 2D. The starting point of the critical distance for breaking the Re–S bond is 0.788 nm, which is comparable to the numerical critical elastic strain of 12% (0.738 nm for $ReS_2$ case) for that direction (table S1). It indicates that the fracture along the [010] direction is driven only by perfect unzipping of Re–S bonds without plastic deformation and even out-of-plane distortion (Fig. S5).

Previous studies reported that the other 2D TMDs (*e.g.*, $MoS_2$, $WS_2$) accompanies reconstructed crack tips with emanated dislocation (*12-14*). Fig. 2F exhibits successive AR-STEM images (Fig. 2F) of [010] cracking for 58s. To check the propagation process, we use an existing hole defect as a reference (marked by a yellow arrow). The straight edges in the [010] cracking are formed by successively unzipping the Re–S bonds. Simultaneously, atomic sharpness of crack tip and perfect lattices in front of the crack head are maintained without lattice reconstruction, dislocation emission, and blunting process, similar to ceramic materials (*e.g.*, Si, $Al_2O_3$) (*31*). To summarize, these phenomena suggest that anisotropic 2D $ReS_2$ exhibits ultimate pure brittleness along the [010] direction.

The cleavages along the [110] direction (white line in Fig. 1A) are also commonly observed after [010] cracks in exfoliated $ReS_2$ flakes, because, as previously mentioned, [110] cracking has a similar ultimate strength and slightly higher critical strain compared to the [010] cracking. AR-STEM image in Fig. S7 shows the atomic-scale crack tip propagating along the [110] direction. The [110] cracking entails continuous unzipping of the Re–S and Re–Re bridge bonds. Similar to the [010] cracking described above, the [110] cracking does not involve dislocation emission and plastic deformation in the vicinity of the crack tip, exhibiting the brittleness of ReS$_2$ along this direction. The [110] cracking is associated with covalent Re–Re bridge bonds, but the low ultimate tensile strength and low critical strain along the [110] might be due to the large spacings between Re$_4$ clusters along this direction (~ 0.30 nm), and the longer bonding length of Re–Re bridge bonds (~ 289 pm) than the Re–Re cluster bonds (~277 pm) (fig. S8). Therefore, the [110] cleavage results in the another most frequently observed cleavage direction after the [010] direction.

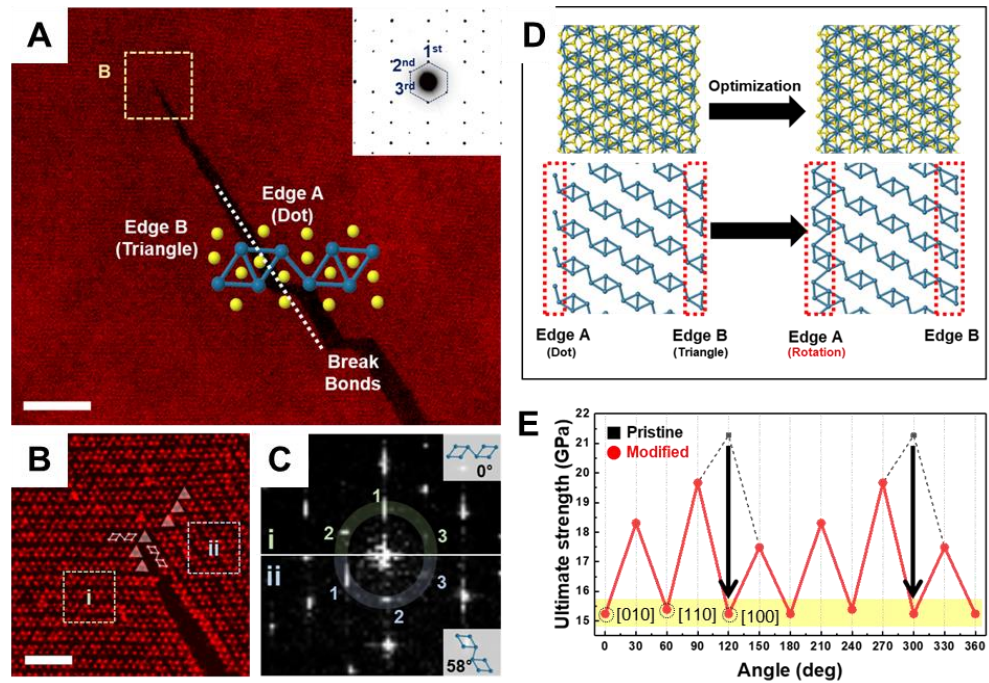

**Figure 3. Crack propagation along the [100] direction (involving breakage of the Re$_4$ cluster).** (**A**) Large field-of-view of the [100] cracking in single-layer ReS$_2$ and corresponding SAED pattern (inset) to determine the exact orientation. Broken Re$_4$ clusters result in edge A (including dot defect) and edge B (including triangle defect). (**B**) Atomic-scale HAADF-STEM image near the [100] crack tip, obtained from yellow region in (A). Left side of crack has lots of triangular defects, whereas right side has rotated Re-chain direction by forming a grain boundary in front of the crack head. (**C**) The corresponding FFT results obtained in regions (i) and (ii) shown in (B). The combination of (B) and (C) confirms that the Re-chain orientation is rotated about 58º in the right side of the crack (including edge A). (**D**) VASP results representing spontaneous Re-chain rotation at edge A. (**E**) Pristine (black dashed line with square symbol) and modified (red solid line with circle symbol) ultimate strength curve according to the angle between the crack propagation direction and the Re-chain direction. We marked the cracking direction at the corresponding angle; [010], [110], [100] cracking is matched up with 0º, 60º, and 120º, respectively. Scale bar for (A) and (B), 20 nm and 2 nm, respectively.

To illustrate the crack propagation mechanism in the crystallographic direction involving strong Re–Re covalent bonding, an atomic-scale investigation of crack propagation along the [100] direction (*i.e.*, diamond-shaped $Re_4$ cluster disassemble) was performed. Figure 3A represents the [100] cracks accompanying the cleavage of two strong Re–Re covalent bonds. The dissociation of $Re_4$ clusters by crack propagation along the [100] direction (white dashed line in Fig. 3A) produces Edge A (one Re atom extracted from the $Re_4$ cluster) and Edge B (triangle-shaped $Re_3$ cluster). Atomic-resolution image of the crack tip (Fig. 3B) magnified in the dashed box region of Fig. 3A shows lots of triangular defects in front of the crack tip and at Edge B (left-side of the crack in Fig. 3B), which are indicated by white triangles superimposed on the image. Interestingly, on the other hand, the $ReS_2$ orientation of right-side of the crack (area (ii) in Fig. 3B,C) is rotated 58º compared to the original $ReS_2$ orientation (inset SAED pattern in Fig. 3A) without flipping over (*i.e.*, keeping "A" structure). To sum up, the orientation of $ReS_2$ on the left-side of the crack containing the triangular defect (edge B) retains its original structure, while the right-side including the dot defect (edge A) rotates approximately 60º clockwise, which is confirmed by the combination of AR-STEM images and the corresponding fast Fourier transform (FFT) analysis (Fig. 3B,C). In addition, the corresponding Re-only atomic models represented in the inset in Fig. 3C show excellent agreement with the experimental AR-STEM image in Fig. 3B. The projection distance between adjacent Re1 (Re4) across the crack along the [100] cracking exhibits the peculiar characters with a reduced distance at the crack tip, indicative of the generation of structural reconstruction (fig. S10). Rationally, when the cleavage occurs along the [100] direction, it is estimated to have a saw-toothed edge including energetically favorable [110] and [010] edges despite increasing edge length, because high energy is required for straight [100] crack (fig. S11C). However, in [100] cracks, straight edges with orientation reconstruction are often observed (fig. S11B,C), and even in the exfoliated flakes, straight and sharp edges are frequently seen rather than saw-toothed edges (fig. S4A, fig. S12A).

Previous studies have reported that the orientation of $ReS_2$ could be easily flipped (*i.e.*, A to A′ orientation, and vice versa), because of the lots of S vacancies irradiated by the electron beam (*32*). Likewise, the tiny displacement of the Re atom readily leads to crystallographic reconstruction due to the high randomness of the low-symmetric $ReS_2$ atomic configuration. To account for the origin of the Re-chain 60º rotation, VASP calculation is conducted using $ReS_2$ nanoribbon models containing Edge A and Edge B. After energy optimization, edge B with triangular defects maintains its original orientation, while, interestingly, edge A with dotted defects is spontaneously reconstructed by forming bonds between adjacent original Re-chains, anchoring the unsaturated Re atoms *via* additional Re–Re bonds. In other words, 60° rotation of the Re-chain direction occurs spontaneously to reduce the energy of the system (~ 35.11 eV, Fig. S13). As supported in Fig. S13, some of the exfoliated flakes cut along the (100) crystallographic plane with edge B configuration are essentially accompanied by rotation in the Re-chain direction. Therefore, it can be concluded that the rotation in the Re-chain direction in $ReS_2$ greatly reduces the ultimate tensile strength and the corresponding cleavage work function, so that it can be sufficiently cleaved along the [100] direction. From the above results, we propose a modified ultimate strength plot (Fig. 3E) and a modified cleavage work function (Fig. S14), reflecting the rotation of the $ReS_2$ orientation.

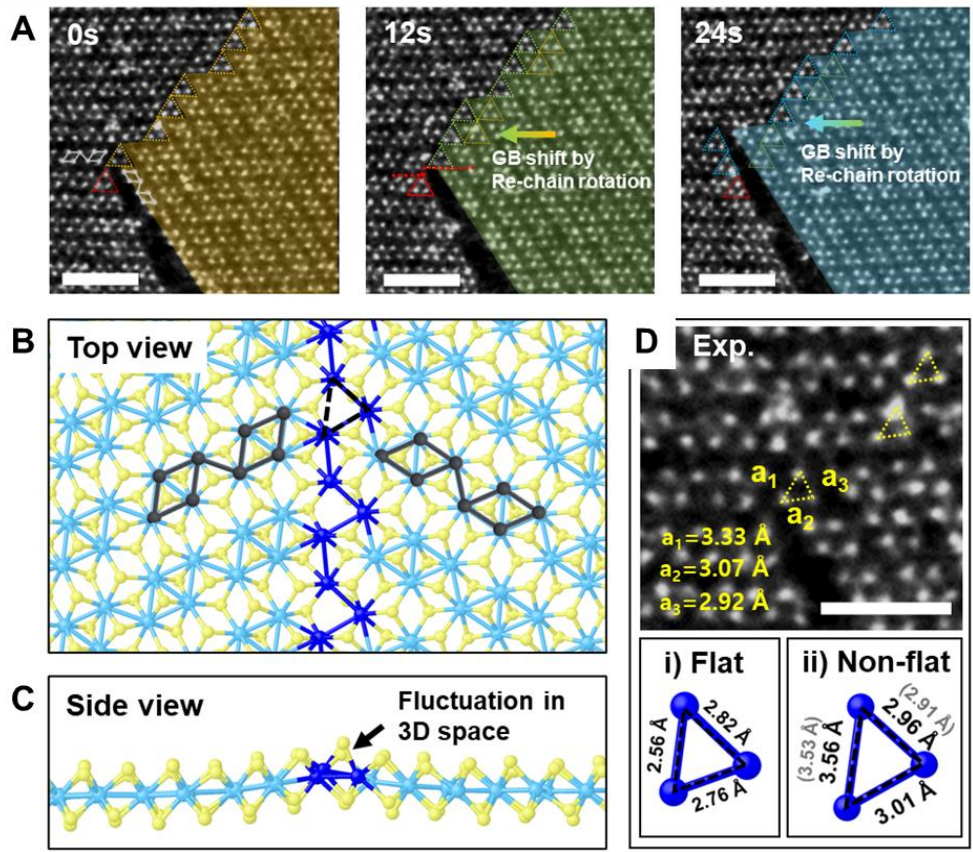

**Figure 4. Grain boundary (GB) migration by the [100] cracking.** (**A**) Successive AR-STEM image series showing *in situ* dynamics of $ReS_2$ crack propagation along the [100] direction. The AR-STEM images show the GB formed in front of the crack tip due to the Re-chain rotation in the right part of the crack. The triangular $Re_3$ clusters that make up the GB shift to the left as the crack propagates along the [100] direction. (**B,C**) Top (B) and side (C) views of the GB atomic model between 60° rotated $ReS_2$ grains. The relaxed structure represents the fluctuations in 3D space along the GB line (denoted by blue atoms). (**D**) The experimental and calculated distances between adjacent Re atoms in the GB component (Re$_3$ cluster). The experimental result is good agreement with the non-flat structure of GB than the flat one. The gray numbers represent the projection distance because the STEM image is a projection image to the 2D plane. The sky blue (blue) and yellow balls represent Re (Re at GB) and S atoms, respectively. Scale bar for (A) and (D), 2 nm.

As forementioned, the [100] cracking spontaneously entails Re-chain rotation on one-side of the crack, thus naturally forming a grain boundary (GB) in front of the crack tip. Figure 4A shows the time-series of AR-STEM images in the vicinity of the crack tip where the crack propagates along the [100] direction to fully understand the anisotropic crack propagation mechanism along [100] direction in atomic scale. Initially (Fig. 4A, 0s), the left side of the crack obviously shows a consistent ReS$_2$ orientation with many triangular defects (red dashed triangles) decomposed by the Re$_4$ clusters, while the opposite side (orange shaded) exhibits Re-chain 58° rotation, resulting in a GB formation in front of crack head (dashed orange triangles). Notably, the Re-chain rotation region (orange-, green- and blue-shaded region in Fig. 4A, 12s and 24s, respectively) is expanded so that the crack-driven GB is shifted upwards compared to 0s. As the (010) plane is converted to the (110) plane by Re-chain 58° rotation, $d$-spacing (0.35 nm for (010), 0.3 nm for (110) plane) is shortened by 14.29%, resulting in the upshift of the right side of crack as in Fig. 4A, 12s (dashed red line).

The [100] crack proceeds accompanying crystallographic reconstruction. At this point, we are wondering why only [100] cracks involve reconstruction, thus we investigated whether the lattice reconstruction occurs in a direction other than the [100] crack. For convenience of explanation, we set the stress direction horizontally and ignore Poisson's ratio for simple comparison. The [010] cracks have an ultimate strain of 12%, the lowest value compared to other directions. Therefore, we examined 12% strained ReS$_2$ structures along different directions (blue atomic models, fig. S15) and a pristine ReS$_2$ structure (orange atomic models, fig. S15) that could have [010] cracks when the strain is applied horizontally. Notably, the strained structure of the 120° case is most similar to the orange-colored pristine ReS$_2$ atomic model. Since structural reconstruction is energetically promoted by structural resemblance, [100] cracks (120° case) are easily reconstructed with a 60° rotation, resulting in [010] cracks. Furthermore, for the case of other direction (the case of 30° depicted in fig. S12B,C), serrated cleavage surfaces consisting (010) and (110) planes alternatively are exhibited, while straight surfaces for the [100] cracks.

Various configurations of GB have been reported in ReS$_2$ because Re-chains are readily reconstructed by small displacements of Re atoms and the formation energy of GB is very low (*24*). In our case, a 60° rotating GB with no flipped structure (here referred to as GB-60°) is observed, which is in good agreement with the atomic model as shown in Fig. 4B. The formation energy of GB-60° is very low, about 1 eV/nm (*24*), indicating an easily formed configuration. After relaxing the GB-containing ReS$_2$ configuration, out-of-plane distortion occurs at GB, which might release the in-plane strain (*33*) as shown in Fig. 4B,C. In order to determine whether the observed GB has a flat or non-flat structure, we carefully investigate the configuration of GB-60°. The distances between adjacent Re atoms of triangular Re$_3$ clusters at GB are measured (> 8 points), and the average distance $a_1$, $a_2$, and $a_3$ are 3.33, 3.07, 2.92 Å, respectively (see Fig. 4D and table S2). Our experimental results are much closer in the non-flat GB structure with fluctuations in 3D space (non-flat atomic structure in Fig. 4D), suggesting that energetically favorable structure by reducing the in-plane strain at GB *via* 3D fluctuation. Atomic models as shown in fig. S16 show the mechanism of GB-60° movement by tiny displacement of Re atoms. As a closing remarks, it can be concluded that the crystallographic rich randomness of ReS$_2$ facilitates the cleavage along the [100] direction; in other words, the lattice reconstruction, which reduces the cleavage work function despite theoretically high resistance to fracture, and the out-of-plane fluctuation to reduce the in-plane strain assist in cutting in the [100] direction as well as [001], [110] direction.

In conclusion, the atomic-scale crack propagation depending on the propagation direction is investigated using the mechanically exfoliated in-plane anisotropic 2D ReS$_2$. The conventional ultimate tensile strength and critical strain of the SS curve exhibits the easiest to cut in the [010] direction (parallel to the Re-chain), followed by the [110] direction, both with large spacings between adjacent Re$_4$ clusters. In atomic-scale view, the [010] and [110] cracking represent atomically sharp crack tips without lattice reconstruction, dislocation emission, and tip blunting, indicative of the genuine brittleness of ReS$_2$. On the other hand, the [100] cracking, energetically unfavorable to break but frequently observed, involves lattice reconstruction. Specifically, crystallographic rotation greatly reduces the cleavage energy, facilitating the cleavage along this direction. The GB fluctuated in 3D space formed by the Re-chain rotation has low formation energy, which enables the flake fracture along the [100] direction. Therefore, our results impart that atomic-scale studies of mechanical dynamics have contributed significantly to a full understanding of the mechanical properties of 2D materials with rich randomness from confined dimension, beyond previous understanding of fracture mechanics.

**Acknowledgements**

This work was supported by Institute for Basic Science (IBS-R019-D1) and the National Research Foundation of Korea (NRF) grant funded by the Korea government (MSIT) (No. 2018R1A2A2A05019598).

**Methods**

**Exfoliation of 2D $ReS_2$ nanosheets.** Thin (single- and bi-layer) $ReS_2$ samples were prepared by mechanical exfoliation from bulk $ReS_2$ single crystals (2D Semiconductors) using adhesive tape. and attached on Si substrates with a 300 nm $SiO_2$ layer. Then, with the help of OM, we selected desirable thin $ReS_2$ samples to investigate crack propagations. In particular mono-layer $ReS_2$ is highly desirable for intuitive interpretation of TEM analysis. To conduct TEM experiments, the flakes on the $SiO_2$/Si substrates were transferred to a holey carbon-supported TEM grid using a wet direct transfer method with the help of isopropyl alcohol (*34*). The wet direct transfer has the advantage of being able to transfer certain flakes.

**TEM imaging and image processing.** TEM measurements were conducted using an aberration-corrected FEI Titan cube G2 60-300 with a monochromator, operated at 80 and 200 kV. SAED pattern was acquired to show the diffraction information of the selected region using the selected-area aperture located in the image plane of the objective lens. To minimize the beam damage on $ReS_2$, STEM analysis was performed by taking images with a short exposure time (8 μs dwell time) with a low electron beam current (55 pA). The acquired images were processed using ImageJ's bandpass filter to correct for long-range non-uniformity of the illumination intensity.

**DFT calculation.** The first-principle calculations are carried out with the Vienna Ab initio Simulation Package (VASP) based on density functional theory (DFT). The exchange–correlation function used is the generalized gradient approximation (GGA) of Perdew-Burke-Ernzerhof (PBE). The electronic plane wave interception energy is set to be 550 eV and for the atomic relaxation and electronic structure calculations of $ReS_2$ unit cells with 24 atoms, a 5×3×2 k-point mesh is used.

The vacuum layer between neighbouring images are set to be larger than 15 Å, which is enough to avoid the interactions between neighbouring images. All the structures are relaxed until the energy differences were converged within $10^{-5}$ eV and the forces of all atoms are less than 0.01 eV/Å.

**References**

1. J. R. Kermode *et al.*, Low-speed fracture instabilities in a brittle crystal. *Nature* **455**, 1224-U1241 (2008).
2. I. H. Lin, R. Thomson, Overview No-47 - Cleavage, Dislocation Emission, and Shielding for Cracks under General Loading. *Acta Metall Mater* **34**, 187-206 (1986).
3. Z. G. Liu, C. Y. Wang, T. Yu, Molecular dynamics simulations of influence of Re on lattice trapping and fracture stress of cracks in Ni. *Comp Mater Sci* **83**, 196-206 (2014).
4. R. Thomson, C. Hsieh, V. Rana, Lattice Trapping of Fracture Cracks. *J Appl Phys* **42**, 3154-& (1971).
5. M. J. Buehler, H. J. Gao, Dynamical fracture instabilities due to local hyperelasticity at crack tips. *Nature* **439**, 307-310 (2006).
6. S. I. Heizler, D. A. Kessler, H. Levine, Mode-I fracture in a nonlinear lattice with viscoelastic forces. *Phys Rev E* **66**,  (2002).
7. Z.-H. J. C.T. Sun, in *Fracture Mechanics*. (Elsevier, 2012), pp. 247-285.
8. K. S. Novoselov *et al.*, Electric field effect in atomically thin carbon films. *Science* **306**, 666-669 (2004).
9. J. H. Warner *et al.*, Dislocation-Driven Deformations in Graphene. *Science* **337**, 209-212 (2012).
10. M. Fujihara *et al.*, Selective Formation of Zigzag Edges in Graphene Cracks. *Acs Nano* **9**, 9027-9033 (2015).
11. X. N. Wang, A. Tabarraei, D. E. Spearot, Fracture mechanics of monolayer molybdenum disulfide. *Nanotechnology* **26**,  (2015).
12. S. S. Wang *et al.*, Atomically Sharp Crack Tips in Monolayer MoS2 and Their Enhanced Toughness by Vacancy Defects. *Acs Nano* **10**, 9831-9839 (2016).
13. B. M. Wang *et al.*, Role of sulphur atoms on stress relaxation and crack propagation in monolayer MoS2. *Nanotechnology* **28**,  (2017).
14. T. H. Ly, J. Zhao, M. O. Cichocka, L. J. Li, Y. H. Lee, Dynamical observations on the crack tip zone and stress corrosion of two-dimensional MoS2. *Nat Commun* **8**,  (2017).
15. G. S. Jung *et al.*, Interlocking Friction Governs the Mechanical Fracture of Bilayer MoS2. *Acs Nano* **12**, 3600-3608 (2018).
16. G. S. Jung *et al.*, Anisotropic Fracture Dynamics Due to Local Lattice Distortions. *Acs Nano* **13**, 5693-5702 (2019).
17. X. Ling *et al.*, Anisotropic Electron-Photon and Electron-Phonon Interactions in Black Phosphorus. *Nano Lett* **16**, 2260-2267 (2016).
18. G. Qiu *et al.*, Observation of Optical and Electrical In-Plane Anisotropy in High-Mobility Few-Layer ZrTe5. *Nano Lett* **16**, 7364-7369 (2016).
19. L. Li *et al.*, Strong In-Plane Anisotropies of Optical and Electrical Response in Layered Dimetal Chalcogenide. *Acs Nano* **11**, 10264-10272 (2017).
20. S. X. Huang *et al.*, In-Plane Optical Anisotropy of Layered Gallium Telluride. *Acs Nano* **10**, 8964-8972 (2016).
21. L. Li *et al.*, Highly In-Plane Anisotropic 2D GeAs2 for Polarization-Sensitive Photodetection. *Adv Mater* **30**,  (2018).

22. W. Wen *et al.*, Anisotropic Spectroscopy and Electrical Properties of 2D ReS2(1-x)Se2x Alloys with Distorted 1T Structure. *Small* **13**, (2017).
23. X. Y. Chen *et al.*, Pristine edge structures of T "-phase transition metal dichalcogenides (ReSe2, ReS2) atomic layers. *Nanoscale* **12**, 17005-17012 (2020).
24. X. B. Li *et al.*, Nanoassembly Growth Model for Subdomain and Grain Boundary Formation in 1T ' Layered ReS2. *Adv Funct Mater* **29**, (2019).
25. S. L. Jiang *et al.*, Direct synthesis and in situ characterization of monolayer parallelogrammic rhenium diselenide on gold foil. *Commun Chem* **1**, (2018).
26. M. Rahman, K. Davey, S. Z. Qiao, Advent of 2D Rhenium Disulfide (ReS2): Fundamentals to Applications. *Adv Funct Mater* **27**, (2017).
27. H. F. Wang *et al.*, Cleavage tendency of anisotropic two-dimensional materials: ReX2 (X = S,Se) and WTe2. *Phys Rev B* **96**, (2017).
28. Y. Choi *et al.*, Complete determination of the crystallographic orientation of ReX2 (X = S, Se) by polarized Raman spectroscopy. *Nanoscale Horiz* **5**, 308-315 (2020).
29. Y. Guo *et al.*, Distinctive in-Plane Cleavage Behaviors of Two-Dimensional Layered Materials. *Acs Nano* **10**, 8980-8988 (2016).
30. D. A. Chenet *et al.*, In-Plane Anisotropy in Mono- and Few-Layer ReS2 Probed by Raman Spectroscopy and Scanning Transmission Electron Microscopy. *Nano Lett* **15**, 5667-5672 (2015).
31. D. R. Clarke, K. T. Faber, Fracture of Ceramics and Glasses. *J Phys Chem Solids* **48**, 1115-1157 (1987).
32. Y. C. Lin *et al.*, Single-Layer ReS2: Two-Dimensional Semiconductor with Tunable In-Plane Anisotropy. *Acs Nano* **9**, 11249-11257 (2015).
33. J. H. Kim *et al.*, Interface-Driven Partial Dislocation Formation in 2D Heterostructures. *Adv Mater* **31**, (2019).
34. W. H. Lin *et al.*, A Direct and Polymer-Free Method for Transferring Graphene Grown by Chemical Vapor Deposition to Any Substrate. *Acs Nano* **8**, 1784-1791 (2014).